# Solving intractable chemical problems by tensor decomposition

Nina Glaser and Markus Reiher *

*Abstract:* Many complex chemical problems encoded in terms of physics-based models become computationally intractable for traditional numerical approaches due to their unfavourable scaling with increasing molecular size. Tensor decomposition techniques can overcome such challenges by decomposing unattainably large numerical representations of chemical problems into smaller, tractable ones. In the first two decades of this century, algorithms based on such tensor factorizations have become state-of-the-art methods in various branches of computational chemistry, ranging from molecular quantum dynamics to electronic structure theory and machine learning. Here, we consider the role that tensor decomposition schemes have played in expanding the scope of computational chemistry. We relate some of the most prominent methods to their common underlying tensor network formalism, providing a unified perspective on leading tensor-based approaches in chemistry and materials science.

**Keywords:** Compression · Machine Learning · Tensor Decomposition · Tensor Networks

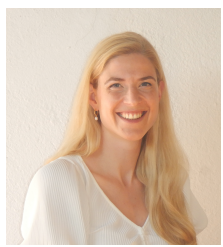

*Nina Glaser* studied Interdisciplinary Natural Sciences at ETH Zürich, Switzerland with a focus on computational physics and chemistry. After a research stay at the University of Colorado Boulder, USA, where she worked on quantum Monte Carlo methods in the group of Prof. Sandeep Sharma, she returned to ETH Zurich to pursue a PhD in the Theoretical Chemistry group of Prof. Markus Reiher. Her current research focus is on the development of novel tensor-based wave function methods for vibrational and electronic structures to solve time-independent and time-dependent molecular many-body problems.

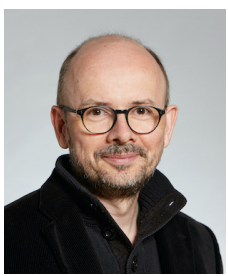

*Markus Reiher* has been a professor of theoretical chemistry at ETH Zurich since 2006. His research focuses on theoretical chemistry in the broadest sense and ranges from relativistic quantum chemistry to the development of new electron-correlation theories and smart algorithms for the autonomous and interactive exploration of complex chemical reaction networks. He applies machine learning as a glue to combine data from different physical models. Recently, he has developed a keen interest in quantum computing after his 2016 work with Microsoft Quantum on nitrogenase as the first real-world application for this new computing paradigm in the molecular and materials sciences.

---

* Correspondence: Prof. Dr. M. Reiher, E-mail: mreiher@ethz.ch, Dept. of Chemistry and Applied Biosciences, ETH Zürich, CH-8093 Zürich

## 1. Introduction

The goal of computational chemistry is to describe the properties and reactivity of molecular systems accurately, yet efficiently, with computational models in order to gain insights into complex chemical processes.[1] Depending on the problem at hand, the involved particles are treated classically, fully quantum mechanically, with hybrid quantum-classical schemes, or with data-driven approaches. While the theoretical foundations of both classical and quantum mechanics are well established, the development of efficient computational methods to calculate stationary molecular properties or to simulate the evolution of complex many-body systems over time remains an active field of research. In the past decades, tremendous efforts were made to push the limits of numerical calculations to larger and more complex chemical systems, fueled by the constantly increasing computational power available and the emergence of novel algorithms. A particularly prevalent, albeit often not directly apparent, common ground of many state-of-the-art approaches in computational chemistry is the utilization of powerful tensor decomposition schemes. Tensor decomposition-based approaches enable the systematic factorization of complex high-dimensional numerical representations of molecular features into many, interconnected, lower-dimensional tensors, which essentially allows one to solve several small problems in *lieu* of one intractably large one. In this context, tensors can be understood as any kind of multidimensional array of (real or complex) numbers, with their dimensionality being defined by the number of indices required to specify a tensor element. For instance, a vector is a one-dimensional tensor, whereas a matrix corresponds to a two-dimensional tensor. In chemistry, tensors often appear within computational models as a consequence of the numerical description of the entities involved (such as electrons, atomic nuclei, atoms, functional groups, and so forth) and their interactions with one another. Due to the unfavorable scaling of the tensor dimension with system size, numerical approaches based on tensor decomposition schemes have emerged as a particularly promising family of computational methods to tackle chemically challenging systems.

In Section 2 of this article, we consider the foundations of tensor decomposition schemes and review one of the most common factorization techniques, the singular value decomposition, before we present some of the most prominent tensor decomposition approaches in computational chemistry in Section 3. We conclude in Section 4 by providing a unified perspective and by discussing the future of these methods in the context of emerging computational developments.

## 2. Tensor Decomposition Techniques

Tensor decomposition schemes have a long history in numerical mathematics and linear algebra.[2] Over the past two centuries, various techniques to factorize matrices have been developed, and some have also been generalized to higher-order tensors.[3] As any high-dimensional tensor can be reshaped into a two-dimensional matrix by an appropriate grouping of indices,

$$\mathbf{T}_{i_1,\ldots,i_k,i_l,\ldots,i_N} \equiv \mathbf{M}_{(i_1,\ldots,i_k),(i_l,\ldots,i_N)} \equiv \mathbf{M}_{\alpha\beta} \quad (1)$$

any matrix factorization technique can effectively also be applied to reshaped higher-order tensors.

One of the most commonly applied matrix factorizations is the singular value decomposition (SVD), which can be viewed as a generalized eigendecomposition applicable to arbitrary matrices. SVD factorizes any real or complex $m \times n$ matrix $\mathbf{M}$ into a product of three matrices,

$$\mathbf{M} = \mathbf{U}\mathbf{\Sigma}\mathbf{V}^* \quad (2)$$

where $\mathbf{U}$ is a $m \times m$ unitary matrix and $\mathbf{V}^*$ is a $n \times n$ unitary matrix in its conjugate transpose form. The matrix $\mathbf{\Sigma}$ is a $m \times n$ rectangular diagonal matrix with non-negative real entries, the so-called singular values, on the diagonal. If the matrix exhibits a high degree of degenerateness, it has many singular values that are equal to zero. The number of singular values that are not zero corresponds to the rank of the matrix.

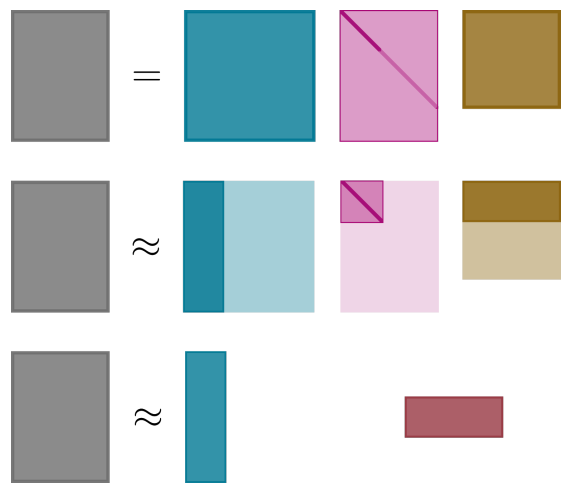

Fig. 1. Top row: An original matrix (gray) is decomposed by SVD into two unitary matrices (blue and brown) and a diagonal matrix (purple). By disregarding all singular values which are zero or negligibly small (middle row), the matrix factorization can be compressed (bottom row).

As many large matrices encountered in chemical modeling do not have full rank, they can be efficiently compressed by only retaining the non-zero singular values, and disregarding the redundant part of the factorization, as visualized in Fig. 1. In numerical applications, matrices may have numerous singular values which are very close to zero and can be deemed negligible, hence effectively truncating the factorization. The compressed matrix obtained by a truncated SVD is the best approximation with a fixed rank that can be obtained for a given matrix as measured by the Frobenius norm.[4] This very powerful property makes the SVD one of the most commonly applied matrix factorization techniques,[2] with routine applications including linear least-squares fitting problems, image compression (see Fig. 2 for an example), tensor-based wave function methods in quantum chemistry (see Sec. 3.2), and machine learning approaches (see Sec. 3.4).

Numerous other tensor decomposition techniques have been developed, among them most notably QR decomposition, Cholesky factorization, LU decomposition, canonical polyadic decomposition, and (hierarchical) Tucker decomposition. For an in-depth description of these techniques, we refer the reader to Refs. [3] and [5]. All of the above-mentioned decomposition schemes are routinely applied in various contexts in computational chemistry. In the following, we review some of the most prominent tensor decomposition-based methods currently developed to push the limits of computational approaches in solving complex chemical problems, ranging from interaction tensor factorization for quantum chemical calculations (see Sec. 3.1.) to molecular quantum dynamics simulations (see Sec. 3.3).

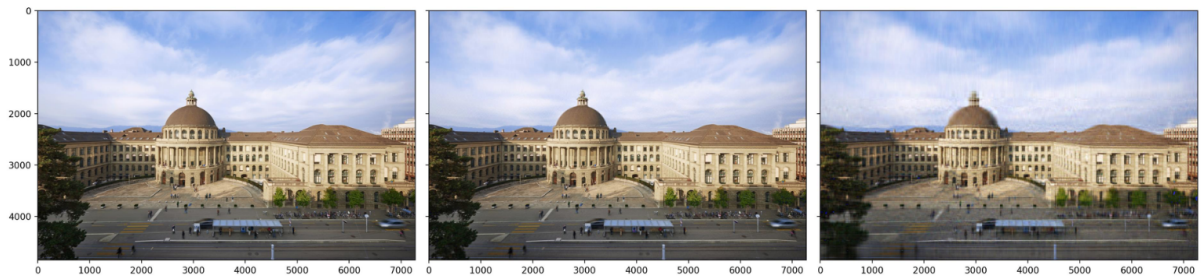

Fig. 2 Image compression with SVD. From left to right: original image, reconstructed image with 5% of the singular values, and with 1% of them. Images can be conveniently expressed as 3-dimensional tensors, with the pixels distributed along the x- and y-axes, and the color encoding (such as RGB) in the third dimension. Image copyright: ETH Zürich / Gian Marco Castelberg

## 3. Tensor Decomposition in Chemistry

### 3.1 *Efficient Encoding of Interactions*

One particularly common task encountered during the computational characterization of molecular many-body systems is the evaluation of interactions between many particles. In classical simulations, the calculation of electrostatic interactions between pairs of atoms (or coarse-grained pseudo-particles) is often the most time-consuming step, as a straightforward evaluation of all pairwise interactions scales with the system size squared.[6] In quantum chemical applications, the computational scaling of the pairwise Coulomb interactions is even steeper, as most calculations require the evaluation of four-index interaction integrals. Not only the generation, but also the manipulation and storage of these four-index tensors pose a great challenge due to their $N^4$ scaling, with $N$ being the number of basis functions (i.e., orbitals) chosen to represent the particles of a molecular system. Since already small molecules easily require a few dozen and more basis functions for a reliable description of the relevant degrees of freedom, efficient schemes to compress the interaction tensors are needed.[7] In the following, we will limit our discussion to the example of two-electron integrals, but many of the techniques mentioned below can also be extended to other multi-index interaction tensors.

A two-electron integral, which is a contribution to the electrostatic energy (consider that the square of two orbitals is a density and the product of two densities over the distance is the integrand of a Poisson integral), can be written in Hartree atomic units for a given atomic-orbital basis set $\{\phi_i\}_{i=1}^N$ as

$$h_{ijkl} = \int\int \phi_i^*(\mathbf{r}_1)\phi_k^*(\mathbf{r}_2)\frac{1}{r_{12}}\phi_j(\mathbf{r}_1)\phi_l(\mathbf{r}_2)\, d\mathbf{r}_1\, d\mathbf{r}_2 \quad (3)$$

where $\mathbf{r}_1$ and $\mathbf{r}_2$ denote the electronic coordinates of two interacting electrons, and $r_{12}$ is the distance between them. All integrals $h_{ijkl}$ can be stored in a four-index tensor, which we abbreviate in chemical notation as $(ij|kl)$. A routinely applied two-electron integral tensor factorization scheme is the resolution of the identity (RI) technique,[8] sometimes also referred to as density fitting. The basic idea of RI is to express the four-index tensor $(ij|kl)$ in terms of (at most) three-index quantities

$$(ij|kl) \approx \sum_\gamma^{N_\gamma} A_{ij}^\gamma B_{kl}^\gamma \quad (4)$$

by inserting a resolution of the identity of $\gamma$, spanned by an auxiliary basis set which is optimized for an efficient approximation of the full integrals. Straightforwardly applying such an RI decomposition would, however, presuppose knowledge of the full four-index tensor, and therefore require the costly evaluation of all two-electron integrals. In order to avoid the full construction of the four-index tensor in the first place, the integrals are directly evaluated in the factorized form, commonly by employing an approximation such as

$$(ij|kl) \approx \sum_{\mu\nu} C_{ij}^\mu V_{\mu\nu} C_{kl}^\nu \quad (5)$$

where $C_{ij}^\nu$ and $C_{kl}^\mu$ are the three-index expansion coefficients of the auxiliary basis set, and $V_{\mu\nu}$ are the two-index electron repulsion integrals over the pre-optimized auxiliary basis functions. Hence, for the calculation of the four-index integral tensor in the RI approximation, only up to three-center integrals are required, of which there are much less than in the original set of integrals containing even four-center integrals (the centers denote the positions of the atomic nuclei in the molecule). Because of its computational efficiency, the RI approximation is routinely applied in quantum chemistry programs to accelerate electronic structure calculations such as density functional theory (DFT).[8]

For applications where the RI approximation is not applicable (e.g., because auxiliary basis functions have not been optimized), a more general Cholesky decomposition (CD) can be applied.[9,10] In CD, a positive semidefinite matrix $\mathbf{M}$ is decomposed as $\mathbf{M} = \mathbf{L}\mathbf{L}^\top$, where $\mathbf{L}$ is a lower triangular matrix and $\mathbf{L}^\top$ is its transpose. Hence, the four-index integral tensor can be approximated with controllable accuracy by

$$(ij|kl) \approx \sum_{n=1}^{N_L} \mathbf{L}_{ij}^{(n)} \mathbf{L}_{kl}^{(n)\top} \quad (6)$$

where $\mathbf{L}^{(n)}$ is the $n$-th Cholesky vector. For many molecular systems, the number of Cholesky vectors required for adequately expressing the two-electron integrals only grows as $N_L \sim N \log N$,[11] therefore effectively reducing the number of terms in the expansion. Directly exploiting the factorized CD form allows for obviating zero or negligibly small integral terms without having to calculate the entire interaction tensor as not all elements of the decomposed matrices are needed to construct the Cholesky vectors $\mathbf{L}^{(n)}$. Such a decomposed form of the two-electron integrals gives rise to what is known as the singly-factorized electronic Hamiltonian.

For large molecular systems, the CD factorization might not be sufficient to compress the integral tensor to a tractable size, such as, for instance, in quantum computation applications with limited resource availability.[12] In such cases, one option is to employ a double factorization of the integrals by further decomposition of the singly factorized CD tensor of Eq. (6). As $\mathbf{L}$ is usually sparse for large molecular systems, it can be further compressed, for instance, with a truncated singular vector decomposition for each single Cholesky vector as

$$\mathbf{L}^{(n)} \approx \sum_m^{N_m} \lambda_m^{(n)} \vec{R}_m^{(n)} \left(\vec{R}_m^{(n)}\right)^\top \quad (7)$$

This results in a doubly-factorized form of the four-index integral tensor, in which this tensor is compactly represented in terms of low-rank vectors,[11] allowing for efficient storage and handling of the two-electron interaction terms.

### 3.2 Electronic Structure Calculations
Electronic structure calculations are central to computational chemistry, as the accurate determination of the electronic energy is crucial for the prediction of molecular properties and the modeling of chemical processes on a molecular level.[13] As a common starting point, most established wave function methods employ an expansion of the many-electron wave function $|\Psi\rangle$ in terms of antisymmetrized products of molecular orbitals,

$$|\Psi\rangle = \sum_{\sigma_1,\ldots,\sigma_L}^{N_{\sigma_1},\ldots,N_{\sigma_L}} c_{\sigma_1,\ldots,\sigma_L} |\sigma_1\rangle \otimes \cdots \otimes |\sigma_L\rangle = \sum_\sigma c_\sigma |\sigma\rangle \quad (8)$$

where $\sigma_i$ is the occupation of the $i$-th molecular orbital $\phi_i$, and $c_\sigma$ is the expansion coefficient that measures the contribution of an electronic configuration $|\sigma\rangle$ to the electronic state $|\Psi\rangle$. If unrestricted, this expansion is known as the full configuration interaction (FCI) wave function and can be viewed as a generalization of the linear combination of atomic orbitals (LCAO) concept to the total electronic wave function. Unfortunately, the number of many-electron basis states $|\sigma\rangle$ grows exponentially with molecule size. Therefore, an exact solution of the electronic structure problem in the space spanned by the electronic configurations of the FCI *ansatz* is only feasible for small molecules, drastically limiting the applicability of this direct approach. For large molecules, efficient methods that capture the relevant quantum effects while lowering the computational costs are needed. Processes, where quantum correlations cannot be reliably approximated (*e.g.,* by Kohn-Sham DFT), such as chemical reactions catalyzed by transition metal compounds, require a sophisticated quantum mechanical treatment for an adequate description of the physical mechanisms.

In the past two decades, the density matrix renormalization group (DMRG) algorithm[14] emerged as a powerful high-accuracy method for efficient calculations of strongly correlated quantum systems in chemistry.[15] While DMRG was originally developed for the simulation of one-dimensional spin lattice models, it has later been extended to quantum chemical applications to capture electron correlation effects in molecular systems.[16,17] Although at first, only pseudo-linear molecules were considered suitable targets for formal reasons, it was then shown that DMRG can be applied to multi-configurational molecular problems in practice.[15] Whereas DMRG was initially not introduced as a tensor-based wave function method, it was eventually realized that the entire algorithm can be elegantly reformulated in terms of tensor factorizations.[18–20] This tensor network formulation of DMRG demonstrates the efficient compression of complex many-body wave functions by factorizing the FCI tensor according to

$$|\Psi\rangle = \sum_\sigma c_\sigma |\sigma\rangle = \sum_\sigma \mathbf{M}^{\sigma_1} \mathbf{M}^{\sigma_2} \cdots \mathbf{M}^{\sigma_L} |\sigma\rangle \quad (9)$$

where the CI coefficients $c_\sigma$ are now expressed as products of $N$ matrices $\mathbf{M}^{\sigma_i}$ (with the first and last, i.e., $\mathbf{M}^{\sigma_1}$ and $\mathbf{M}^{\sigma_L}$, are row and column matrices (vectors), respectively). The resulting wave function

*ansatz* is known as a matrix product state (MPS), and is also referred to as a tensor train in mathematics.

The decomposition of the FCI tensor containing all CI

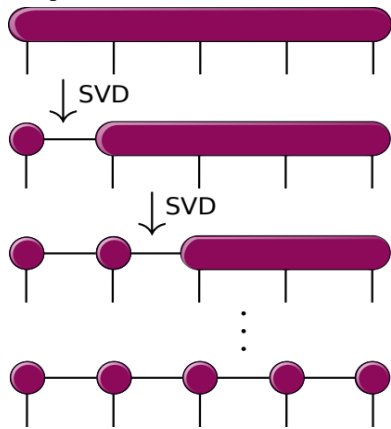

Fig. 3. Tensor decomposition of the FCI tensor (top) into an MPS (bottom) by sequential SVDs. In the tensor diagrams, the filled shapes represent the tensors, with the adjoined lines visualizing their respective indices.

coefficients into MPS form by sequential SVDs is illustrated in Fig. 3. The key advantage of the MPS representation is that it allows for a very effective compression of the full wave function by simply limiting the maximum size of the matrices $\mathbf{M}^{\sigma_i}$ to dimension $m \times m$ through retaining only the $m$ largest singular values in each step. This ensures that the most significant electronic configurations are included in the MPS whereas negligible contributions are truncated, resulting in compact yet accurate multi-configurational wave functions. The approximation degree can be conveniently controlled by adapting the truncation threshold.

To compactly represent molecular wave functions numerically, one usually works directly with the factorized tensor *ansatz*, thus bypassing the explicit construction of the full-dimensional tensors, which would be unfeasible anyways. Hence, numerical algorithms are required which allow for an efficient calculation of the tensor factorization without having access to the full-dimensional tensor, as the FCI wave function of complex molecular systems is generally not known. For MPSs the tensor parameters can be determined with the DMRG algorithm, which solves the time-independent Schrödinger equation through a variational optimization of the wave function. As the direct solution of the full-dimensional eigenvalue problem is computationally intractable for large systems, DMRG exploits the factorized *ansatz* by optimizing one tensor after the other. In an iterative (so-called sweeping) procedure, all tensors of the factorization are updated repeatedly until convergence is reached. During this sequential MPS optimization, many small eigenvalue problems need to be solved numerically, rather than one big one. Interestingly, the small eigenvalue problems are typically too large for direct diagonalization approaches to yield the electronic energy, and hence, subspace iteration methods such as Davidson diagonalization[22,23] need to be employed, which represent another set of tensor approximation methods. Because of its favorable scaling with molecular size, the DMRG algorithm allows one to efficiently calculate ground- and excited-state wave functions of electronic, vibrational and vibronic problems, as well as molecular properties that are otherwise very difficult to handle.[21,24] Finally, we note that the MPS format has then also been discovered in the analytic construction of multi-configurational wave function *ansätze* in the multifacet graphically contracted function method.[25]

### 3.3 *Molecular Quantum Dynamics*
Simultaneously to DMRG emerging as a state-of-the-art electronic structure method for strongly correlated systems, the multilayer multi-configurational time-dependent Hartree (ML-MCTDH) algorithm[26–28] was developed in the field of molecular quantum dynamics to enable high-accuracy nuclear dynamics simulations with many degrees of freedom. While these two developments occurred largely independently of one another and (originally) with very different target applications, it was later realized that the two approaches are related as they can both be reformulated based on tensor network states.[29] Similarly to DMRG, the molecular wave function is expanded in terms of single-particle functions in ML-MCTDH, but instead of a linear MPS factorization, the wave function *ansatz* corresponds to a hierarchical Tucker decomposition, which is visualized in Fig. 4. The resulting tensor factorization is also known as a tree tensor network state (TTNS), and it allows for an efficient encoding of hierarchical interaction patterns.

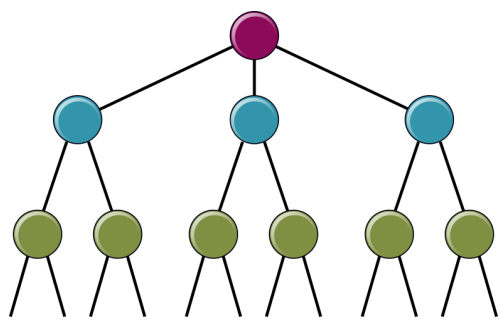

Fig. 4. In this TTNS, the differently colored tensors highlight the different layers in the hierarchical Tucker decomposition employed in ML-MCDTH.

Similar to the MPS *ansatz*, the TTNS *ansatz* allows for a compression of molecular wave functions by limiting the size of the multi-dimensional tensors in the network to some maximum size $n_{\kappa_i}$ for the $\kappa_i$-th index. In analogy to the MPS factorization in DMRG, the construction of the full-dimensional tensor is avoided also in ML-MCTDH, and instead, the hierarchical Tucker decomposition is directly employed as an *ansatz* for the nuclear wave function. However, in contrast to standard DMRG as applied to molecular structure problems, ML-MCDTH is commonly employed to simulate quantum dynamics, and thus the tensors change over time as the wave function evolves. Hence, the coefficients are not determined through an energy minimization scheme as in time-independent DMRG calculations, but instead, the tensors are propagated in time according to equations of motion derived from the time-dependent variational principle.

However, a similar strategy to solve the time-dependent Schrödinger equation has also been applied to MPSs. The MPS wave function is then propagated in time with the time-dependent DMRG (TD-DMRG) algorithm based on the time-dependent variational principle.[30,31] In a TD-DMRG calculation, each tensor is updated in a sequential fashion similarly to the time-independent DMRG algorithm, until the entire MPS wavefunction has been propagated over a given time step. By applying suitable time-propagation algorithms to tensor network states, such as TTNSs or MPSs, high-accuracy simulations of both nuclear and electronic dynamics in molecules can be performed.[32,33,34]

### 3.4 *Machine Learning Approaches*

In addition to molecular simulations which are directly based on (quantum) mechanical principles, a new class of computational approaches is becoming increasingly prevalent within chemistry, namely, data-driven methods based on machine learning (ML) models.[35–37] With the advent of sophisticated ML algorithms and the increase in available chemical data, ML has opened up a complementary route to traditional approaches. By employing powerful statistical models, ML enables an efficient characterization of systems which are too large or there are too many to be explicitly simulated mechanically with the given computational resources. There are two key areas of application where ML is routinely exploited to expand the limits of traditional computational chemistry. First, ML can further the understanding of chemical processes by extracting patterns in complex, high-dimensional data. Second, ML can interpolate chemical data to efficiently predict properties of molecular systems based on data-driven models. Many of the ML approaches commonly applied to these two classes of problems in computational chemistry can be viewed as tensor factorization-based techniques employing statistically optimized tensor decompositions of large, complex problems.

A well-known example of an ML method routinely applied to gain insights into complex, high-dimensional data is principal component analysis (PCA).[38] PCA is a linear dimensionality reduction technique based on projecting high-dimensional data sets onto a low-dimensional space (usually two or three dimensions for visualization purposes). The new coordinates, the so-called principal axes, span the space with the largest variation in the data. While PCA is often formulated as an eigendecomposition of the covariance matrix of the data set, it can also be understood as the direct application of a SVD (see Sec. 2) to the data matrix itself. In practice, a PCA via SVD is numerically more efficient in most cases. The singular values obtained from SVD are proportional to the variance of the data associated with each eigenvector. Hence, the largest singular values indicate the most important coordinate axes, and correspondingly, the associated right singular eigenvectors represent the principal components of a data set. One key difference of PCA compared to most of the other tensor-factorization-based approaches referred to in this paper is that PCA is employed to reduce the dimensionality of a chemical data set *a posteriori*: the full data tensor is known and only later decomposed. Therefore, the factorized tensors can be calculated straightforwardly by SVD, without the need of applying involved algorithms to determine the tensor coefficients. This makes PCA an easily applicable tool for gaining insight into high-dimensional chemical data in cheminformatics and chemometrics, and can be applied, for instance, for the design and planning of chemical libraries.[39]

A large variety of ML-based approaches has been developed for a second key application in computational chemistry, namely, the efficient prediction of chemical properties of molecular systems. For the modeling of chemical processes and materials, a prevalent target problem is the evaluation of the potential energy for a given atomistic structure. Potential energy evaluations are a frequent task in various applications, ranging from reaction network explorations to molecular dynamics simulations.[40] While classical molecular dynamics simulations based on parametrized force fields allow for cost-efficient evaluations of the potential energy and the resulting atomic forces, classical force fields fail to fully capture the underlying interactions of electrons and atomic nuclei of the molecular system. A proper quantum mechanical description of the potential can be achieved with *ab initio* molecular dynamics, where the potential energy is obtained directly from an electronic structure calculation (in Born-Oppenheimer molecular dynamics) or by propagation of orbitals (as in Car-Parrinello molecular dynamics) at each simulation time step. However, this approach is severely limited with respect to system size and simulation timescales due to the cost of the electronic-structure part, even if one restricts them to methods such as DFT. Fortunately, significant advances have been made in this respect with the development of machine learning potentials (MLPs).[41,42] These potentials aim to preserve the high accuracy of first-principles methods while lowering the computational demands to those of classical force fields. MLPs are commonly based on very flexible mathematical expressions which are parametrized by optimizing the predictions for a given reference data set of molecular structures and corresponding *ab initio* energies and forces. At the highest level, all MLPs can be viewed as sophisticated interpolation schemes on the training data to predict the potential energies of similar molecular structures. Various types of MLPs have been developed, with a particularly frequently applied family of potentials, namely the high-dimensional neural network potentials (HDNNPs) which are based on feed-forward neural networks (NNs).[43] Roughly speaking, NNs encode a hierarchy of concepts relating the input, which in this case are molecular descriptors, to the target output, which here corresponds to the potential energy. An illustration of such a feed-forward NN is given in Fig. 5.

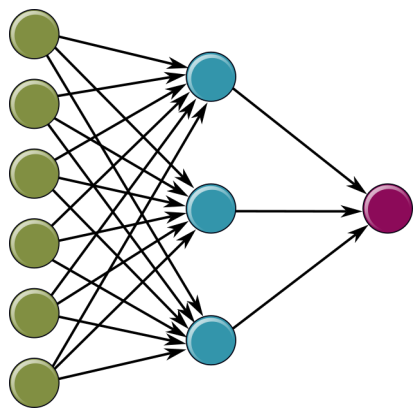

Fig. 5. Feed-forward neural network. Data is mapped from the input layer (green) to the target output quantity (purple) through parametrized connections (arrows) via hidden layers (blue).

Mathematically, the collection of interconnected nodes can be expressed in terms of tensors, which contain all the parameters of the different connections between two layers and user-defined analytical functions describing the relationship from one layer to the next. Such networks of parameter tensors combined with suitable nonlinear transformations result in very expressive models to encode the mapping between a given input to the target output. As for almost all previously introduced tensor-decomposition-based approaches, the key numerical challenge also for HDNNPs is the determination of the tensor coefficients, as they cannot be calculated through a straightforward tensor decomposition. While for the examples in Sections 3.1 to 3.3 the tensor entries were determined through equations arising from the underlying physical principles, the parameters of feed-forward neural networks are learned by optimizing the coefficients of each connection during training of the ML model. For this purpose, a cost (or loss) function has to be defined which provides a quantitative measure of how well the model performs for a given reference example. The NN parameters are then optimized with respect to this cost function by employing a suitable optimization procedure, usually some type of backpropagation algorithm based on gradient descent,[44,45] until convergence is reached.

While the training process of an MLP is computationally quite expensive, as it requires costly reference data sets and entails numerous optimization iterations, all subsequent evaluations of the potential can be computed very efficiently. To ensure accurate potential energy predictions of molecular structures, MLPs can be combined with uncertainty quantification techniques to monitor the model variance error, for instance by employing an ensemble of models. For molecular structures with an energy or force uncertainty predicted above a given target threshold, the MLP can be improved with further training on such structures by employing active learning or lifelong learning schemes, which allow for a continuous optimization of the model parameters.[46] These properties make MLPs a powerful concept to combine the efficiency of classical force fields with the accuracy of quantum chemical calculations.

## 4. Conclusions

As the field of computational chemistry aims to gain insight into chemical processes through computational models of molecular systems, tensors naturally arise in various contexts from the numerical representation of the involved particles and the quantification of their interactions. For larger many-body systems, a straightforward numerical description can quickly become prohibitive due to the unfavorable scaling of the tensor dimensionality with system size. By systematically factorizing complex high-dimensional problems into many interconnected lower-dimensional ones with tensor decomposition schemes, previously intractably large chemical systems can be targeted. For this reason, numerical methods based on tensor decompositions are becoming ubiquitous in computational chemistry as a plethora of tensor factorization-based approaches has emerged in the past two decades.

While different tensor decomposition-based approaches are applied in very diverse research fields, they share the underlying assumption that tensor factorizations provide an efficient description of the interacting many-body systems commonly encountered in chemistry. Suitable tensor decomposition schemes often allow for a compact, while accurate representation of molecular systems with their intricate physical interactions governing chemical processes. Depending on the problem at hand, tensor factorizations are leveraged either to efficiently encode the interactions between particles, to compactly express multi-configurational many-body wave functions, or to flexibly parametrize complex relations between molecular descriptors and chemical properties. The coefficients of the tensor factorization scheme employed are commonly determined by some sort of functional minimization, which can be based either directly on physical principles, such as the variational principle, or on user-defined cost functions measuring the expressiveness of a given tensor parametrization, as is often the case for machine learning models. A key component of tensor-based methods are therefore the efficient numerical procedures employed to compute the tensor coefficients, such as the DMRG algorithm for molecular structure calculations or adaptive gradient-based backpropagation algorithms for training HDNNPs. For applications where dimensionality reduction is the main goal, such as PCA, the coefficient determination becomes trivial, as the parameters can be calculated with a straightforward decomposition of the full data tensor.

The development of various tensor decomposition-based approaches has often occurred simultaneously while also largely independently in different research areas. Different methods were originally intended for very distinct target applications and have initially been formulated with a specific focus. One particular example for this observation is the emergence of DMRG in solid-state physics for the calculation of spin chains simultaneously to MCTDH for molecular quantum dynamics. As both of these methods were then elaborated and their range of application was extended,[31,47] it was realized that they can both be reformulated based on tensor network states for a flexible representation of a large variety of many-body wave functions. Through the common foundation of efficient tensor factorizations, conceptual and algorithmic connections between the two methods became apparent, allowing for a synergistic advancement of these methods. Also in other research contexts there are many cross-disciplinary developments that are continuously advancing tensor-based approaches, particularly in regard to the prevalence of ML in chemical and materials sciences, which is fueled further by the development of specialized hardware specifically optimized for tensor operations, such as graphical processing units and tensor processing units.

Within the next decade, entirely new approaches for simulating molecular systems and materials may emerge with the advent of quantum computing for real-world chemistry problems.[48] Also in this area, tensor factorization-based techniques will be of great importance, for instance for the calculation of high-quality guiding states or for the design of compact circuits to model complex chemical Hamiltonians. Hence, tensor decomposition-based methods will continue to play a crucial role in tackling intractable chemical problems.

*Acknowledgements*

N.G. is grateful to Metrohm and the Swiss Chemical Society for the award for the best oral presentation in computational chemistry at the SCS fall meeting 2023. The authors gratefully acknowledge the financial support from the Swiss National Science Foundation through grant No. 200021_219616.

Received: 29.01.2024